\documentclass[runningheads]{llncs}
\usepackage[T1]{fontenc}
\usepackage{graphicx,verbatim}
\usepackage{hyperref}
\usepackage{color}

\usepackage{lmodern}           
\usepackage{fix-cm}            
\usepackage[scr=rsfs]{mathalpha} 
\usepackage{multirow}%
\usepackage{amsmath,amssymb,amsfonts}%
\usepackage{mathrsfs}%
\usepackage[title]{appendix}%
\usepackage{xcolor}%
\usepackage{textcomp}%
\usepackage{manyfoot}%
\usepackage{booktabs}%
\usepackage{algorithm}%
\usepackage{algorithmicx}%
\usepackage{algpseudocode}%
\usepackage{listings}%
\usepackage{subcaption}
\usepackage{gensymb}
\usepackage{wrapfig}
\usepackage{multicol}
\usepackage{makecell} 
\usepackage{tabularx}
\usepackage{colortbl}
\usepackage{orcidlink}

\usepackage{soul}
\usepackage{tikz}
\usetikzlibrary{calc} 

\makeatletter
\newif\if@anonymize

\@anonymizefalse  

\if@anonymize
  \newcommand{\highlight@DoHighlight}{
    \fill [outer sep = -15pt, inner sep = 0pt, color=black]
          ($(begin highlight)+(0,8pt)$) rectangle ($(end highlight)+(0,-3pt)$) ;
  }

  \newcommand{\highlight@BeginHighlight}{
    \coordinate (begin highlight) at (0,0) ;
  }

  \newcommand{\highlight@EndHighlight}{
    \coordinate (end highlight) at (0,0) ;
  }

  \newdimen\highlight@previous
  \newdimen\highlight@current
  \newlength{\item@width}

  \DeclareRobustCommand*\anonymize{%
    \SOUL@setup
    \def\SOUL@preamble{%
      \begin{tikzpicture}[overlay, remember picture]
        \highlight@BeginHighlight
        \highlight@EndHighlight
      \end{tikzpicture}%
    }%
    \def\SOUL@postamble{%
      \begin{tikzpicture}[overlay, remember picture]
        \highlight@EndHighlight
        \highlight@DoHighlight
      \end{tikzpicture}%
    }%
    \def\SOUL@everyhyphen{%
      \discretionary{%
        \SOUL@setkern\SOUL@hyphkern
        \SOUL@sethyphenchar
        \tikz[overlay, remember picture] \highlight@EndHighlight ;%
      }{%
      }{%
        \SOUL@setkern\SOUL@charkern
      }%
    }%
    \def\SOUL@everyexhyphen##1{%
      \SOUL@setkern\SOUL@hyphkern
      \settowidth{\item@width}{##1}%
      \makebox[\item@width]{}%
      \discretionary{%
        \tikz[overlay, remember picture] \highlight@EndHighlight ;%
      }{%
      }{%
        \SOUL@setkern\SOUL@charkern
      }%
    }%
    \def\SOUL@everysyllable{%
      \begin{tikzpicture}[overlay, remember picture]
        \path let \p0 = (begin highlight), \p1 = (0,0) in \pgfextra
          \global\highlight@previous=\y0
          \global\highlight@current =\y1
        \endpgfextra (0,0) ;
        \ifdim\highlight@current < \highlight@previous
          \highlight@DoHighlight
          \highlight@BeginHighlight
        \fi
      \end{tikzpicture}%
      \settowidth{\item@width}{\the\SOUL@syllable}%
      \makebox[\item@width]{}%
      \tikz[overlay, remember picture] \highlight@EndHighlight ;%
    }%
    \SOUL@
  }
\else
  \newcommand{\anonymize}[1]{#1}
\fi
\makeatother

\begin{document}
%
\title{Direct vascular territory segmentation on cerebral digital subtraction angiography}

\author{P. Matthijs van der Sluijs\inst{1}\orcidlink{0000-0002-4934-0933} \and Lotte Strong \inst{1} \and Frank G. te Nijenhuis\inst{1}\orcidlink{0009-0003-1321-5836} \and Sandra Cornelissen\inst{1}\orcidlink{0000-0002-0332-2158} \and Pieter Jan van Doormaal \inst{1}\orcidlink{0000-0002-7398-2864}\and Geert Lycklama {\`a} Nijeholt\inst{2}\orcidlink{0000-0001-6575-9868} 
\and Wim van Zwam\inst{3}\orcidlink{0000-0003-1631-7056}\and Ad van Es \inst{4}\orcidlink{0000-0001-6119-7247} \and Diederik Dippel \inst{1}\orcidlink{0000-0002-9234-3515}\and Aad van der Lugt\inst{1}\orcidlink{0000-0002-6159-2228} \and Danny Ruijters\inst{5}\orcidlink{0000-0002-9931-4047} \and Ruisheng Su\inst{5}\orcidlink{0000-0002-5013-1370} \and Theo van Walsum\inst{1}\orcidlink{0000-0001-8257-7759}} 
\authorrunning{P.M. van der Sluijs \textit{et al.}}
\institute{
Erasmus MC, Doctor Molewaterplein 40, 3015 GD Rotterdam, The Netherlands\\
\and
Haaglanden MC, Den Haag, The Netherlands\\
\and
Maastricht UMC+, Maastricht, The Netherlands\\
\and
Leiden UMC, Leiden, The Netherlands\\
\and
TU Eindhoven, Eindhoven, The Netherlands\\
Correspondence: \email{r.su@tue.nl}
}

\titlerunning{Direct vascular territory segmentation on cerebral DSA}
    
\maketitle              
\begin{abstract}
X-ray digital subtraction angiography (DSA) is frequently used when evaluating minimally invasive medical interventions. DSA predominantly visualizes vessels, and soft tissue anatomy is less visible or invisible in DSA. Visualization of cerebral anatomy could aid physicians during treatment. This study aimed to develop and evaluate a deep learning model to predict vascular territories that are not explicitly visible in DSA imaging acquired during ischemic stroke treatment. We trained an nnUNet model with manually segmented intracranial carotid artery and middle cerebral artery vessel territories on minimal intensity projection DSA acquired during ischemic stroke treatment. We compared the model to a traditional atlas registration model using the Dice similarity coefficient (DSC) and average surface distance (ASD). Additionally, we qualitatively assessed the success rate in both models using an external test. The segmentation model was trained on 1224 acquisitions from 361 patients with ischemic stroke. The segmentation model had a significantly higher DSC (0.96 vs 0.82, p<0.001) and lower ASD compared to the atlas model (13.8 vs 47.3, p<0.001). The success rate of the segmentation model (85\%) was higher compared to the atlas registration model (66\%) in the external test set. A deep learning method for the segmentation of vascular territories without explicit borders on cerebral DSA demonstrated superior accuracy and quality compared to the traditional atlas-based method. This approach has the potential to be applied to other anatomical structures for enhanced visualization during X-ray guided medical procedures. The code is publicly available at \url{https://github.com/RuishengSu/autoTICI}.

\keywords{Digital subtraction angiography \and Deep learning \and Segmentation \and Vessel territory \and Unclear borders \and Ischemic stroke.}

\end{abstract}
\section{Introduction}\label{sec1}
X-ray digital subtraction angiography (DSA) is frequently used when evaluating minimally invasive medical interventions. DSA predominantly visualizes vessels, and soft tissue anatomy is less visible or invisible in DSA. Visualization of cerebral anatomy could aid physicians during treatment. 
With advancements in deep learning, the potential to directly segment soft tissue structures from X-ray images arises, allowing for better anatomical context during the procedure. Our research question is whether these structures without well-defined borders could be directly segmented on X-ray images.
This could be valuable during thrombectomy in acute ischemic stroke, where fluoroscopy and DSA guide the intervention. Bones, instruments, and vessels are well visualized, but the affected brain tissue regions are not. Segmenting and visualizing soft tissue regions could provide critical insights into sub-optimal perfusion and its impact on specific regions or vascular territories in the brain\cite{tolhuisen2022value}.

In the context of thrombectomy, the direct segmentation of vascular territories in cerebral DSA presents challenges. DSA images depict vessels, but the vascular regions do not have clear anatomical boundaries, which hampers direct segmentation,  especially in cases of proximal occlusions, where fewer vessels are opacified during contrast injection. Therefore, manual segmentation is often used~\cite{Reder2024euro,Songjnis-2024-022428}. To address this, segmentation using a registration method of anatomical information from an atlas of healthy patients is typically used\cite{wu2007optimum,aljabar2009multi,bach2015atlas}. 
Atlas registration has also been utilized to map vascular territories, such as those of the anterior cerebral artery (ACA), internal carotid artery (ICA), and middle cerebral artery (MCA), onto EVT DSA images, for which autoTICI~\cite{su2021autotici} is an example. 
However, challenges persist with atlas-based registration approaches due to significant differences between the atlas DSA and clinically acquired DSA. These differences include variations in head scaling and rotation as well as which vessels are visualized, resulting in inaccuracies in over 20\% of single-view registrations and approximately 35\% of multi-view patient-level registrations \cite{vanderSluijs2024}. Addressing this task as a segmentation problem could potentially improve these results.

Therefore, the purpose of this study is to develop and assess a deep learning-based approach to segment regions without clear anatomical boundaries in DSA images. The primary application is automated quantification of reperfusion of the ischemic brain area after thrombectomy \anonymize{using autoTICI}. However, this deep learning method could also be applied to other brain regions.

\section{Methods}
\label{section:methods_reg}


\subsubsection{Model}
For the deep learning framework, we selected no-new-UNet (nnUNet)~\cite{isensee2021nnu,isensee2019automated} version 2, which is regarded as a robust baseline model for medical image segmentation. This model was trained using manually segmented vessel territories for the internal carotid artery (ICA) and middle cerebral artery (MCA), derived from \anonymize{autoTICI atlases} for pre- and post-EVT DSA MinIPs\cite{su2021autotici}. The atlases were manually adjusted for size, rotation, and scaling. All annotations were performed in MeVisLab 3.0.2~\cite{heckel2009object}, using an in-house developed tool.


\begin{figure}[!h]
\centering
\includegraphics[width=1\textwidth]{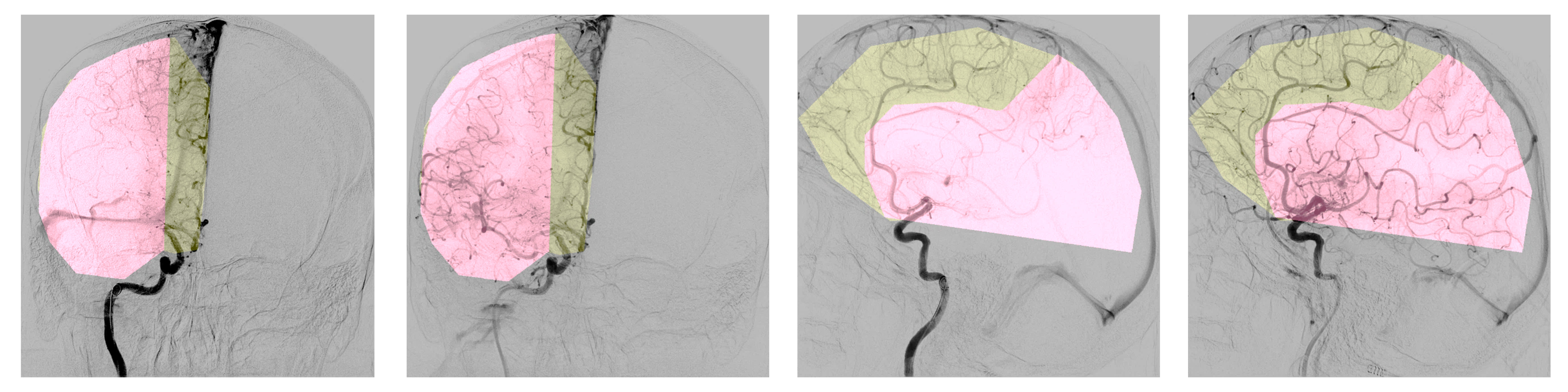}
\caption{From left to right MinIPs of the AP-preEVT, AP-postEVT, Lateral-preEVT, Lateral-postEVT. The yellow area depicts the anterior cerebral artery (ACA) vascular territory. The red area depicts the MCA vascular territory.}
\label{fig:reference_standards}
\end{figure}

The pre-processing involved generating 2D MinIPs from the included DSA sequences ($\text{2D} + \text{time}$) resized to $1024\times1024$ pixels. Three labels were assigned to segment the background, MCA, and ACA territories. The ACA mask was generated by subtracting the MCA from the ICA, as ICA territory is divided into ACA and MCA territory (Fig.~\ref{fig:reference_standards}).
Morphological operations, including erosion, connected component analysis, and dilation, were applied to remove residual ICA lines from the ACA mask.

\subsubsection{Model training}
The dataset was split into $10\%$ for testing and $90\%$ for training, with further subdivision into an $80/20$ training-validation split. A patient-based stratification ensured that pre- and post-EVT images from the same patients were kept in the same split, while maintaining balance in occlusion locations (ICA, M1 and M2). 
Training was conducted on the NVIDIA GeForce RTX 2080 Ti with 11 GB of memory and NVIDIA A40 GPUs with 48 GB of memory within the Erasmus MC GPU cluster, alongside a local system with an NVIDIA GeForce RTX 2080Ti GPU. 
We employed five-fold cross-validation with 1000 epochs, and used model ensembling to average the weights of the models. Initially, separate models were trained for the AP and lateral views, however, a single model trained on both AP and lateral views showed non-statistical significant different performance compared to the separate models (data not shown). This single model for AP and lateral view was preferred for simplicity reasons.
The predicted segmentations were post-processed using the same morphological operations - erosion, connected components analysis, and dilation - as during pre-processing. The ICA mask was reconstructed by combining the ACA and MCA labels.

\subsubsection{Performance metrics}
The model performance was evaluated using Dice Similarity Coefficient (DSC), Jaccard Index (JI), Average Surface Distance (ASD), and the symmetrical Hausdorff Distance (HD). We also evaluated models on their computational time. Depending on data normality, performance metrics were reported either as means with 95\% Confidence Intervals (CIs) or medians with Inter Quartile Ranges (IQRs). We compared the models' performance using either a two-sided paired Student's $t$-test or a Wilcoxon signed rank test, depending on the normality of the data distribution.

\section{Experiments \&\ Results}
\subsection{Data}
We used data from the \anonymize{MR CLEAN Registry} 
a prospective multicenter registry of consecutive patients treated with EVT across \anonymize{19 different hospitals} in the \anonymize{Netherlands between March 2014 and December 2018}. We selected patients with an ICA, M1, or M2 occlusion, who achieved successful reperfusion (eTICI $\geq$2B), and had available DSAs with all vascular phases— arterial, capillary, and venous phase, as well as a non-contrast phase. The dataset comprises DSAs acquired both prior to thrombectomy (pre-EVT), where the anatomical boundary is more obscured due to the occlusion, and after thrombectomy (post-EVT), where the removal of the occlusion makes the anatomical boundary more discernible. The post-EVT of the patients who achieved successful reperfusion was necessary to ensure that the territory that was to be annotated was visible, so that the invisible territory in the pre-EVT could be annotated precisely. From the initial 2756 stroke patients who underwent successful EVT, 2395 patients were excluded based on the following imaging criteria: both AP and lateral views of the post-EVT DSA were required to include all vascular phases of contrast passage (non-contrast, arterial, capillary, and venous), without substantial subtraction artifact due to patient motion, substantial restricted brain field of view or oblique acquisitions. These criteria were applied to ensure reliable manual segmentation of vascular territory. Ultimately, 1,224 eligible acquisitions from 361 patients were included in the analysis, consisting of 651 AP views and 573 lateral views from 504 pre-EVT and 720 post-EVT DSAs. For training, the dataset was split into 90\% for training and 10\% for test, maintaining balance in occlusion locations and ensuring that pre- and post-EVT DSA from the same patients were kept in the same split.
We compared the success rate of the model with \anonymize{the autoTICI-atlas method} developed in a previous study. For this specific analysis, we included the same patients, only omitting those who overlapped with the training and test sets, which resulted in the inclusion of 564 out of 660 patients from the previous study~\cite{vanderSluijs2024}.

\subsection{Experiments}
Three experiments were conducted. The first was training a segmentation model on both AP and lateral view, and comparing it to the existing atlas registration method from \cite{su2021autotici,vanderSluijs2024}. The second experiment was DSA phase dependency by testing the performance on phase-specific MinIPs (arterial, capillary, venous, and non-contrast). The third experiment was to compare the success rate of the segmentations against the atlas method in pre-EVT and post-EVT DSA images.


\subsubsection{Model performance}
We benchmarked the performance of the proposed segmentation model against the atlas registration method evaluating the segmentations using the performance metrics.
The segmentation model showed significantly better DCS, JI, ASD and HD values compared to the atlas registration method. For the ICA territory, the segmentation model achieved a DSC of 0.96 (IQR 0.9-0.97) compared to 0.8 (IQR 0.6-0.8) for the atlas method (p $<$ 0.001), while the ASD for the ICA was 14 (IQR 10-21) vs 47 (IQR 30-67) pixels, p $<$ 0.001)(Table \ref{tab:ap_lat_combined}) The computational time of the segmentation model was significantly lower compared to the atlas registration method (segmentation model: 4s [95\% CI 3.9-4.5] vs atlas 141s [95\%CI 88-223], p $<$ 0.001).\par

\begin{table}[!t]
    \centering
    \caption{\label{tab:ap_lat_combined} The performance metrics of ICA and MCA territory segmentations. The atlas registration method is shown for comparison. Values in the table represent medians and Inter Quartile Ranges.}
\resizebox{\textwidth}{!}{
\begin{tabular}{lllll}

& \textbf{DSC}  & \textbf{JI} & \textbf{ASD} & \textbf{HD} \\ \hline

& \multicolumn{4}{c}{\textbf{ICA territory}} \\
\textbf{Model} 
& 0.96 {[}0.94-0.97{]} & 0.92 {[}0.89-0.93{]} & 14 {[}10-21{]}  & 43 {[}34-59{]}            \\
\textbf{Atlas}    
& 0.82 {[}0.62-0.80{]}***  & 0.70 {[}0.62-0.80{]}***  & 47 {[}30-67{]}***   & 119 {[}84-175{]}***       \\ \hline
& \multicolumn{4}{c}{\textbf{MCA territory}} \\
\textbf{Model} 
& 0.94 {[}0.92-0.96{]} & 0.89 {[}0.85-0.92{]} & 42 {[}35-47{]} & 45 {[}35-98{]}            \\
\textbf{Atlas}    
& 0.78 {[}0.70-0.85{]}*** & 0.64 {[}0.54-0.74{]}*** & 69 {[}50-88{]}*** & 119 {[}88-204{]}*** \\ \hline
    \end{tabular}}
\noindent\parbox{\textwidth}{%
\footnotesize{\textit{Metrics are stated as median [IQR], ASD and HD are in pixels, *** p$\leq$ 0.001 Abbreviations, ASD, Average Surface Distance; DSC, Dice Similarity Coefficient; HD, symmetrical Hausdorff distance; ICA, Internal Carotid Artery; IQR, Inter Quartile Range; JI, Jaccard index; MCA, Middle Cerebral Artery; px, pixels.}}}
\end{table}

\subsubsection{Phase-specific analysis}
\begin{figure} [!ht]
\centering
\includegraphics[width=0.9\textwidth]{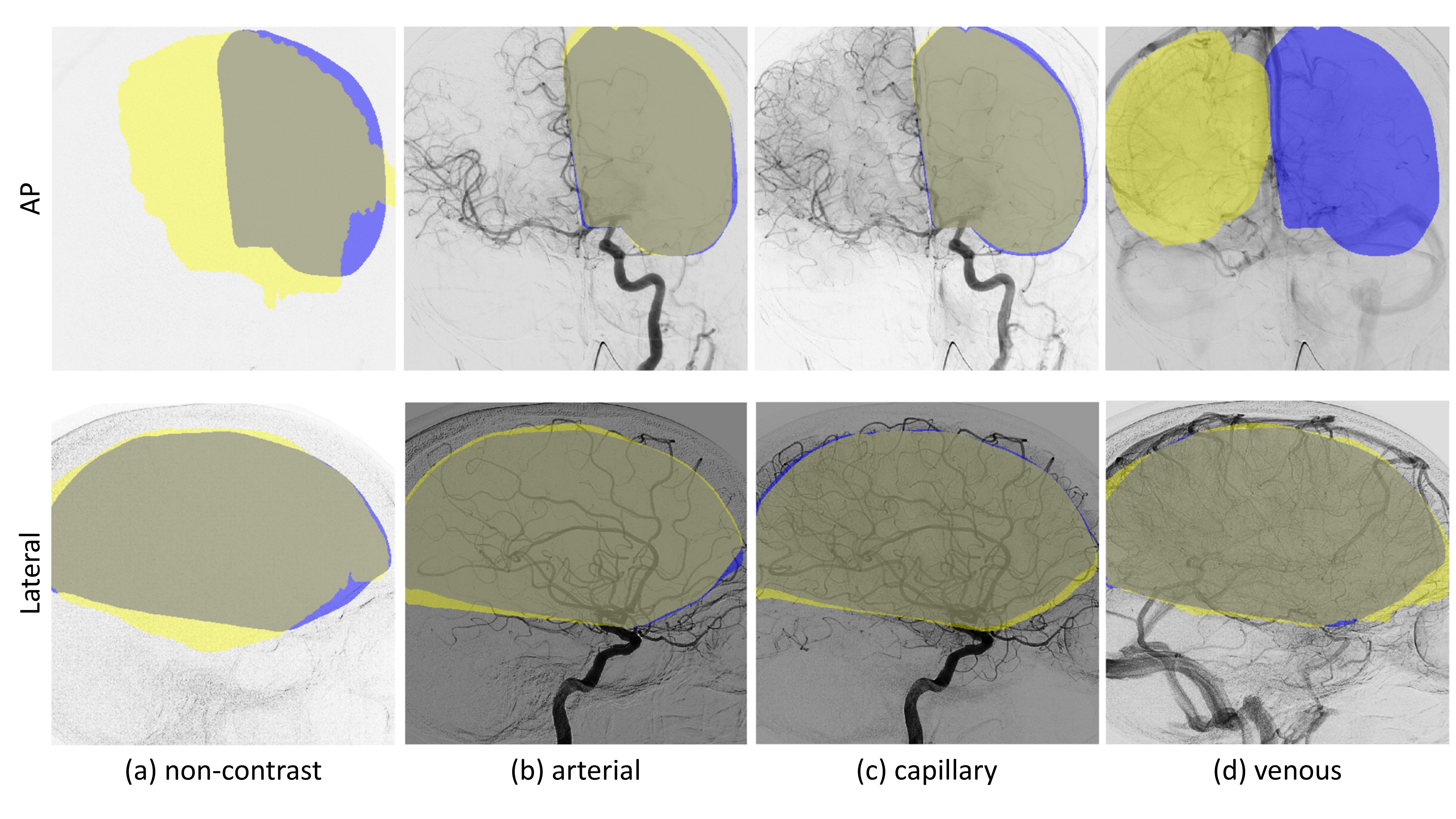}
\caption{\label{fig:split_phases}ICA segmentations for the phase-specific MinIPs of example test set images in the AP and lateral view, depicted in yellow. Blue segmentations represent the segmentations of the model on the full-phase DSA.
\textit{Abbreviations, AP, Anteroposterior; ICA, Internal Carotid Artery}}
\end{figure}

The segmentation model was trained using MinIPs that included all vascular phases— arterial, capillary, venous, and non-contrast frames. To assess which phases influenced the model predictions, we calculated the DSC, JI, ASD, and HD between the segmentations predicted on the full-phase MinIP with those predicted on MinIPs created for each phase. The MinIPs were split into phase-specific MinIPs using the deep-learning phase selection model proposed earlier~\cite{su2021autotici}. We analyzed the non-contrast phase as well to assess whether the model uses the skull outline present in frames due to motion artifacts.
Phase-specific segmentation results for capillary phases were similar to those obtained from DSAs containing all phases. Both the segmentations from arterial and venous MinIPs had slightly different segmentations compared to the full-phase MinIP segmentation (Table \ref{tab:phase_specific}).
The segmentations in the non-contrast frames exhibited significantly lower DSC and JI and higher ASD and HD values when compared to the capillary-phase MinIP. Fig.~\ref{fig:split_phases} visualizes examples of phase-specific segmentations compared against the segmentations produced by the model on the same full-phase DSA. The segmentations in the arterial and capillary phases show high similarity with the full-phase segmentation. In contrast, venous phase predictions displayed incorrect anatomical placements, and non-contrast segmentations were highly erroneous, especially in AP views. Lateral views exhibited slightly more accurate predictions, although with uneven MCA contours in venous and non-contrast phases.

\begin{table}[!ht]
    \centering
    \caption{\label{tab:phase_specific} Performance metrics for ICA and MCA territory segmentations predicted by the model for the MinIPs of each vascular phase: arterial, capillary, venous, and non-contrast.} 
\resizebox{\textwidth}{!}{%
\begin{tabular}{lllll}
\hline
& \textbf{DSC} & \textbf{JI} & \textbf{ASD} & \textbf{HD} \\ \hline
& \multicolumn{4}{c}{\textbf{ICA territory}}                \\
\textbf{Arterial}  & 0.97 {[}0.92-0.98{]}   & 0.94 {[}0.92-0.95{]}   & 15 {[}12-23{]}  & 29 {[}23-41{]}   \\
\textbf{Capillary} & \textbf{0.98 {[}0.97-0.98{]}***}& \textbf{0.96 {[}0.95-0.97{]}***}& 18 {[}14-26{]}  & \textbf{24 {[}15-31{]}***}\\
\textbf{Venous}    & 0.97 {[}0.95-0.97{]}   & 0.93 {[}0.90-0.95{]}   & 18 {[}14-26{]}  & 42 {[}25-53{]}   \\
\textbf{Non-contrast} & 0.82 {[}0.58-0.90{]} & 0.70 {[}0.41-0.82{]}  & 51 {[}30-110{]} & 208 {[}86-394{]} \\ \hline
& \multicolumn{4}{c}{\textbf{MCA territory}} \\
\textbf{Arterial}  & 0.96 {[}0.96-0.97{]}   & 0.93 {[}0.91-0.94{]}   & 39 {[}33-45{]}  & 26 {[}18-34{]}  \\
\textbf{Capillary} & \textbf{0.98 {[}0.96-0.98{]}***}& \textbf{0.96 {[}0.93-0.97{]}***}& \textbf{15 {[}11-21{]}***}& \textbf{18 {[}13-29{]}***} \\
\textbf{Venous}    & 0.96 {[}0.94-0.97{]}   & 0.92 {[}0.89-0.94{]}   & 41 {[}37-48{]}   & 32 {[}21-46{]} \\
\textbf{Non-contrast} & 0.83 {[}0.41-0.89{]}& 0.70 {[}0.26-0.81{]}   & 59 {[}45-130{]}  & 147 {[}73-411{]} \\ \hline
\end{tabular}}
\noindent\parbox{\textwidth}{%
\footnotesize{\textit{Metrics are stated as median [IQR], ASD and HD are in pixels, *** p < 0.001. Abbreviations: ASD, Average Surface Distance; DSC, Dice Similarity Coefficient; HD, symmetrical Hausdorff distance; ICA, Internal Carotid Artery; IQR, Inter Quartile Range; JI, Jaccard index; MCA, Middle Cerebral Artery; px, pixels.}}}
\end{table}

\subsubsection{Comparison against traditional method}
We compared the success rate of the segmentations to assess their suitability for implementation in autoTICI. In autoTICI the atlas registration is performed on the post-EVT MinIP, while the pre-EVT MinIP is registered to the post-EVT MinIP. \par
Segmentations were independently scored by two raters using a Likert scale ranging from 0 to 3 for anatomical correctness. A value of 0 indicated a ‘failure’, where the segmentation was located in the incorrect hemisphere and/or was outside the boundaries of the expected anatomical area; 1 was ‘marginal’ indicating that the segmentation was in the correct hemisphere, but substantially (>10\%) outside the boundaries of the anatomical area; 2 indicated `acceptable', meaning the segmentation was minimally outside the expected anatomical area (<10\%), and 3 denoted a perfect segmentation that matched anatomical expectations. Disagreement was overcome by consensus. One rater had 1 year of experience, while the other had 4 years of experience. 
We compared the proportions success rate of paired data with the McNemar test and the unpaired proportions of the occlusion locations with the $\chi^2$-test.\par

\begin{figure} [!t]
\centering
\includegraphics[width=0.95\textwidth]{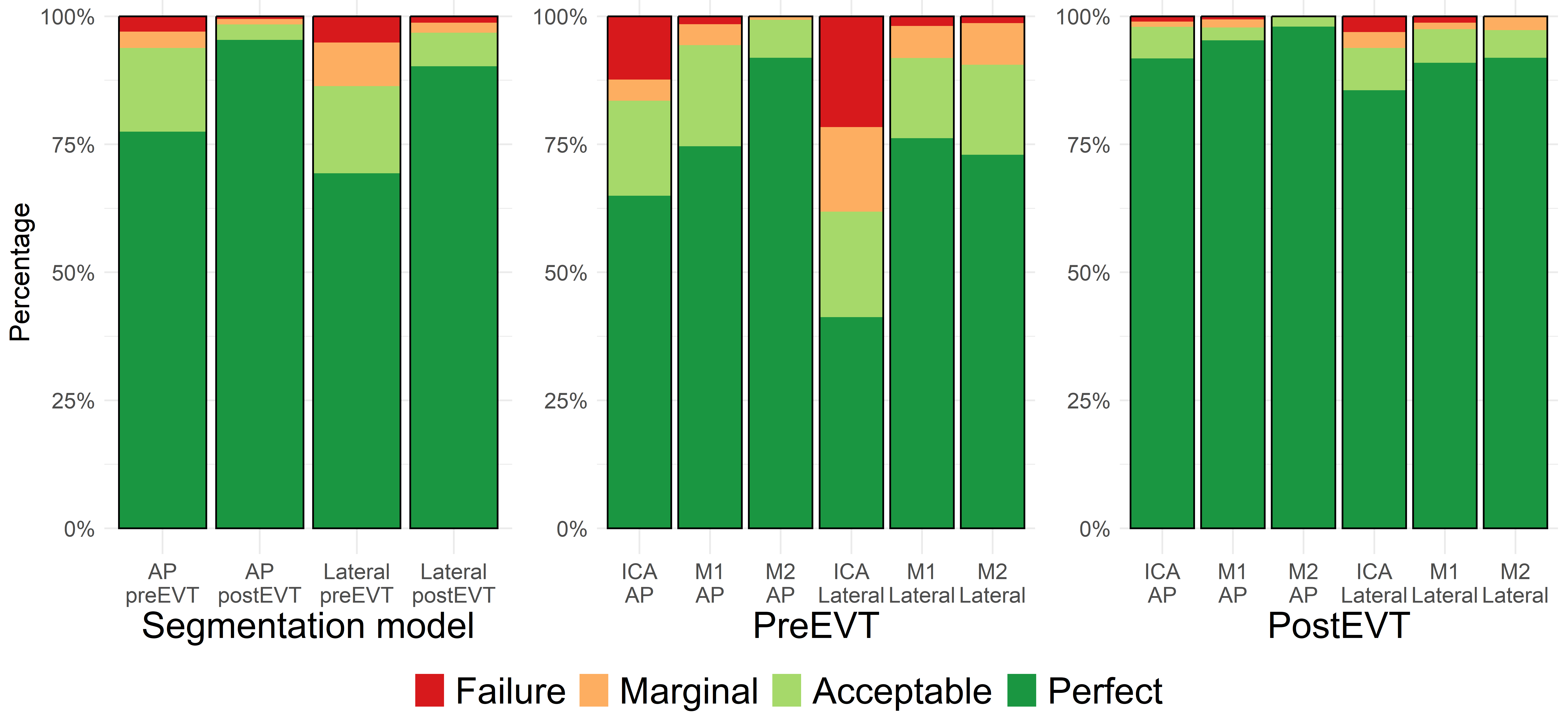}
\caption{\label{fig:likert_segm_pre_post}Distributions (\%) of annotated Likert scores for the segmentation model. Stratified for pre- and post-EVT DSA and occlusion location. 
\textit{Abbreviations; AP, anteroposterior; pre-EVT, pre thrombectomy DSA; post-EVT, post thrombectomy DSA}}
\end{figure}

The segmentation model had a better overall Likert score in the post-EVT compared to the pre-EVT, both in AP and in lateral views (AP success rate pre-EVT 94\% vs post-EVT 98\%, p < 0.001, lateral pre-EVT 86\% vs post-EVT 97\%, p < 0.001, Fig.~\ref{fig:likert_segm_pre_post}). The Likert scores of ICA occlusions in both AP and lateral view were lower compared to the M1 and M2 occlusions (AP success rate: ICA 84\% vs M1 94\%, p = 0.001, ICA 84\% vs M2 99\%, p < 0.001, lateral: ICA 62\% vs M1 92\%, p < 0.001, ICA 62\% vs M2 91\%, p < 0.001, Fig.~\ref{fig:likert_segm_pre_post}).\par

The atlas registration had comparable Likert scores in the post-EVT compared to the pre-EVT, both in AP and in lateral view (AP success rate pre-EVT 86\% vs post-EVT 89\%, p = 0.06, lateral pre-EVT 79\% vs post-EVT 81\%, p = 0.16. The Likert scores of ICA occlusions in both AP and lateral view were comparable in M1 and M2 occlusions (AP success rate: ICA 84\% vs M1 86\%, p = 0.73, ICA 84\% vs M2 90\%, p = 0.2, lateral: ICA 82\% vs M1 78\%, p = 0.46, ICA 84\% vs M2 78\%, p = 0.46.
When considering a per-patient success, where both views in pre- and post-EVT are graded with a Likert score of acceptable or higher, the segmentation model had a higher overall success rate compared to the atlas registration model (80\% vs 66\%, p < 0.001).

\section{Discussion}
In this study we developed a deep learning model to predict vascular regions without explicit borders in DSA imaging of ischemic stroke patients treated with EVT and evaluated against an existing segmentation model based on atlas registration. The proposed model demonstrated superior segmentation performance achieving higher segmentation accuracy. Additionally, it substantially improved the segmentation success rate compared to the atlas registration method in an existing pipeline.
The phase split analysis was conducted to gain insight into which contrast phase the proposed model relied on and showed that the model primarily relied on the capillary phase for accurate segmentation. Segmentation in arterial and venous phases was also effective, likely due to visual residual contrast agent in the arteries during the capillary phase and in the capillaries during the venous phase. In clinical practice, arterial and capillary phases are usually present during DSA for stroke treatment, while venous frames may be absent due to early termination of the acquisition. The non-contrast phase showed worse segmentation compared to the capillary phase, due to limited reliable features in the image. Accurate vascular territory delineation without any visible arteries might not be feasible, even for experienced clinicians. As DSA is used to appreciate the vascular structures, if very few or no vessels are depicted, the might be little use for specific vessel territory assessment.\par 
The success rate for the atlas registration method observed in this study (66\%) was comparable to those previously reported (64\%)\cite{vanderSluijs2024}.  Incorporating the segmentation model would greatly improve the overall success rate of autoTICI to 80\%. As a convenient attribute, the computational time substantially improves to a clinically relevant time from more than 2 minutes to less than 10 seconds. This large difference is probably due to the time-consuming selection of an atlas out of 21 atlases.
Segmentations had a lower success rate in the pre-EVT compared to the post-EVT. This was particularly evident in cases with an ICA occlusion, where the limited number of visible vessels likely contributed to lower performance compared to more distal occlusions. In such proximal ICA occlusions, the brain vasculature is minimally visible, and skull motion artifacts remain, which potentially resembles the non-contrast frames from the phase-split analysis, which corresponded to significantly lower segmentation results. The relatively limited availability of pre-EVT ICA occlusion DSA images may also have contributed to the lower performance, and increasing the size of the dataset could potentially enhance the efficacy. Training a model on specific phases might improve the performance, specifically in ICA occlusions as it could learn the skull motion artifact for a better segmentation. Selecting a certain phase, for example the capillary phase might improve the overall segmentation performance and success rate. However, as DSA is acquired manually there is no assurance that a certain phase is present, thus using all the information from all available phases is a solid option.\par
To our knowledge, this study is the first to apply an automated deep learning-based approach for segmenting vascular territories on cerebral DSA. Previous studies on DSA segmentation focused exclusively on vessels and intracranial aneurysms\cite{zhang2020neural,patel2023evaluating,meng2020multiscale,vepa2022weakly,liu2024dias,neumann2018convolutional,patel2020multi,jin2020fully,patel2020multi,patel2023evaluating,su2024cave}. These studies found mean DSC scores between 0.80 and 0.94, similar to our findings. However, direct comparison is limited, as these studies focused on vessel segmentation, whereas our approach targeted vascular territories. 
Given the performance and success rate of our method, extending this approach to other vascular territories, such as the posterior cerebral artery, or other critical brain areas such as the Wernicke and Broca areas, could provide significant clinical value\cite{brust1976aphasia,tolhuisen2022value}. 
Further research is needed to determine its effectiveness for smaller brain regions, such as the individual components of the basal ganglia.\par
Several limitations should be acknowledged. First, the manual annotation of reference standards could have introduced bias, particularly given the difficulty in visually delineating borders due to over-projection. As such, the results may not entirely reflect the true vascular territories. Future research should aim to reduce this bias by using more objective and accurate reference standard segmentations delineated on CTA and co-registered with DSA \cite{patel2023evaluating}.

\section{Conclusions}
This study proposed a deep learning method for segmenting vascular territories without explicit borders on cerebral DSA. The method is able to produce accurate segmentations and significantly outperforms an atlas registration approach.

\subsubsection{\discintname} The authors have no competing interests to declare that are relevant to the content of this article.

%
%
%
\bibliographystyle{splncs04}
\bibliography{mybibliography}

\newpage
\appendix
\section*{Appendix A}

\begin{figure} [!ht]
\centering
\includegraphics[width=1.0\textwidth]{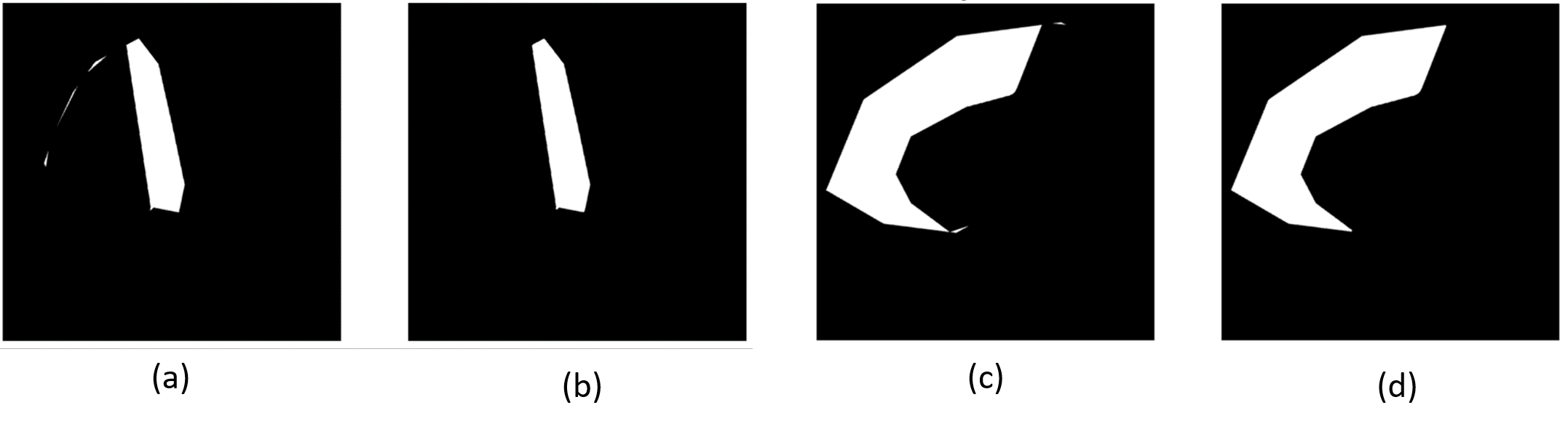}
\caption{\label{fig:preprocessing}Visualization of the morphological operations — erosion, connected component analysis, and dilation for the pre- and post-processing — on the ACA mask (ICA - MCA): (a) ACA mask (AP), with residual ICA mask lines; (b)  ACA mask (AP) after morphological operations; (c) ACA mask (lateral) with residual ICA mask lines; (d) ACA mask (lateral) after morphological operations.}
\end{figure}

\begin{figure} [!ht]
\centering
\includegraphics[width=1.0\textwidth]{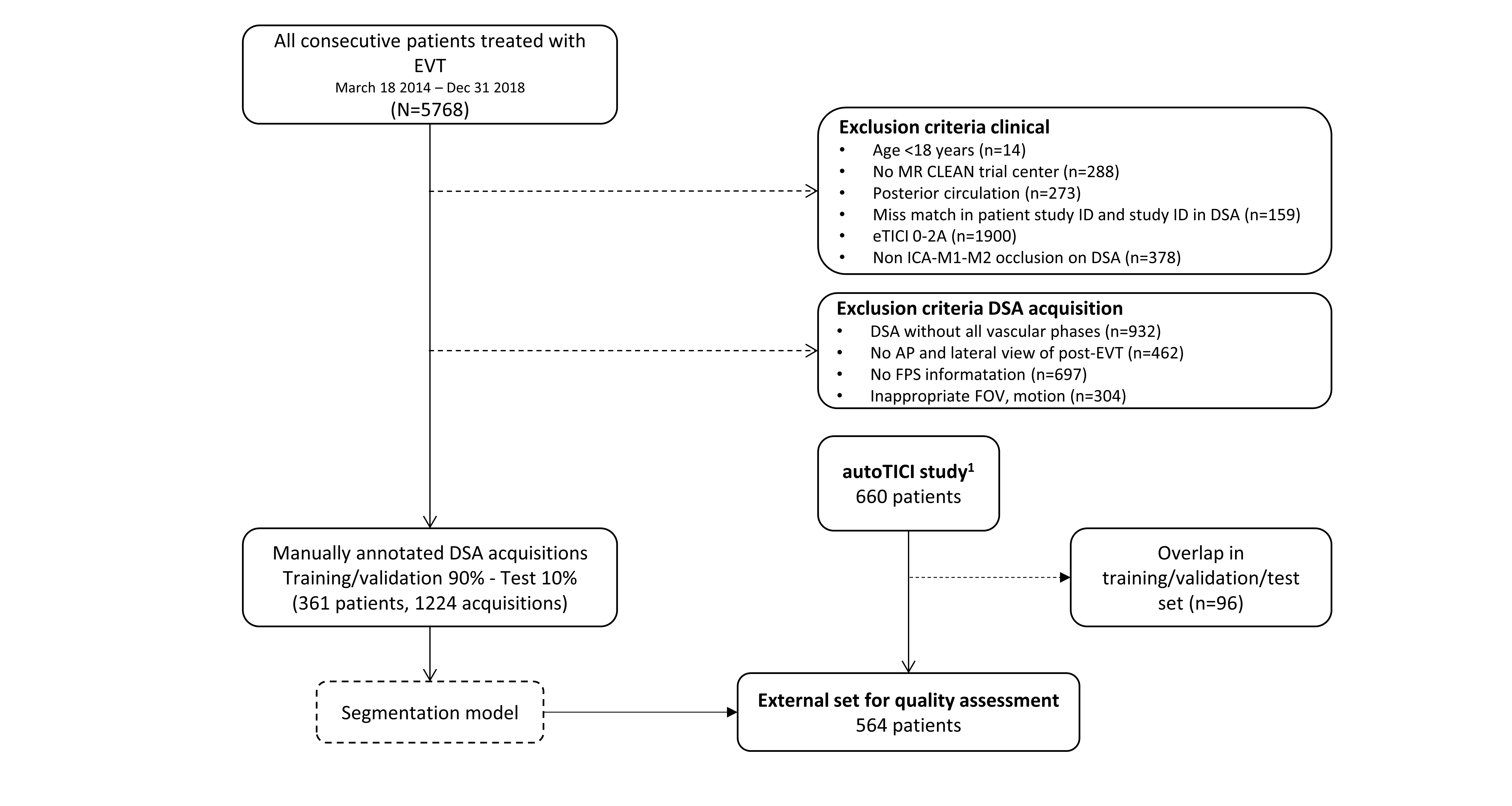}
\caption{\label{fig:Flowchart} Flow chart of the included patients.
\textit{Abbreviations; eTICI, expanded Trombolysis in Cerebral Infarction score}}
\end{figure}


\begin{table}[ht]
\resizebox{\textwidth}{!}{%
\begin{tabular}{|c|c|c|c|c|c|c|c|c|}
\hline
            \rowcolor{lightgray}
 & \multicolumn{4}{c|}{\textbf{Training / Validation}}                                                                   & \multicolumn{4}{c|}{\textbf{Test}}                                                                                    \\
    \rowcolor{lightgray}
        & \multicolumn{2}{c}{\textbf{pre-EVT}}                          & \multicolumn{2}{c|}{\textbf{post-EVT}}                         & \multicolumn{2}{c}{\textbf{pre-EVT}}                          & \multicolumn{2}{c|}{\textbf{post-EVT}}                         \\
  \rowcolor{lightgray}
          & \multicolumn{1}{c}{\textbf{AP}} & \multicolumn{1}{c}{\textbf{Lateral}} & \multicolumn{1}{c}{\textbf{AP}} & \multicolumn{1}{c|}{\textbf{Lateral}} & \multicolumn{1}{c}{\textbf{AP}} & \multicolumn{1}{c}{\textbf{Lateral}} & \multicolumn{1}{c}{\textbf{AP}} & \multicolumn{1}{c|}{\textbf{Lateral}} \\ \hline
\cellcolor{lightgray} \textbf{ICA, n (\%)} & 33 (13)                & 26 (14)                     & 50 (15)                & 49 (15)                     & 6 (20)                  & 3 (13)                      & 7 (20)                 & 8 (24)                      \\ \hline
\cellcolor{lightgray} \textbf{M1, n (\%)}  & 139 (53)               & 98 (52)                     & 175 (54)               & 177 (54)                    & 16 (53)                & 15 (63)                     & 20 (57)                & 17 (52)                     \\ \hline
\cellcolor{lightgray} \textbf{M2, n (\%)}  & 88 (34)                & 66 (35)                     & 100 (31)               & 100 (31)                    & 8 (27)                 & 6 (25)                      & 8 (23)                 & 8 (24)                     \\
\hline
\end{tabular}}
\caption{\label{fig:table_data_stratification} 
\textit{Abbreviations; ICA, Intracranial carotid artery, M1-2, M1 or M2 segment of the middle cerebral artery, pre-EVT, pre thrombectomy DSA; post-EVT, post thrombectomy DSA}}
\end{table}

\begin{figure} [!ht]
\centering
\includegraphics[width=1.0\textwidth]{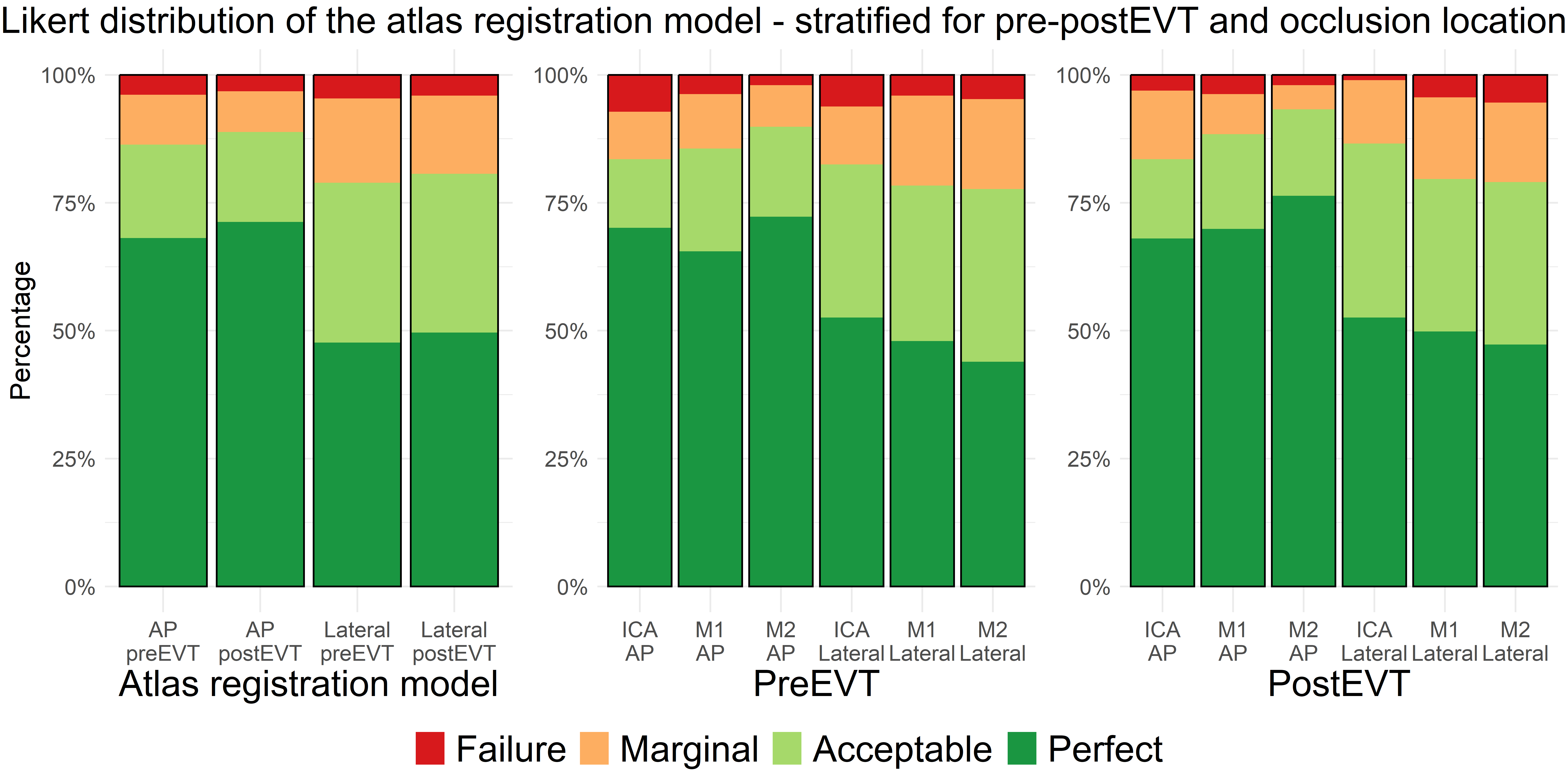}
\caption{\label{fig:likert_atlas_pre_post}Distributions (\%) of annotated Likert scores for the atlas registration model. Stratified for pre- and post-EVT DSA and occlusion location. 
\textit{Abbreviations; pre-EVT, pre thrombectomy DSA; post-EVT, post thrombectomy DSA}}
\end{figure}

\newpage
\section*{Appendix B}
\textbf{MR CLEAN Registry Investigators – group authors}\newline
\textbf{Executive committee}
\newline
Diederik W.J. Dippel$^1$; Aad van der Lugt$^2$; Charles B.L.M. Majoie$^3$; Yvo B.W.E.M. Roos$^4$; Robert J. van Oostenbrugge$^{5,41}$; Wim H. van Zwam$^{6,41}$; Jelis Boiten$^{14}$; Jan Albert Vos$^8$
\newline
\newline
\textbf{Study coordinators}
\newline
Ivo G.H. Jansen$^3$; Maxim J.H.L. Mulder$^1,2$; Robert-Jan B. Goldhoorn$^{5,6,41}$; Kars C.J. Compagne$^2$; Manon Kappelhof$^3$; Josje Brouwer$^4$; Sanne J. den Hartog$^{1,2,40}$; Wouter H. Hinsenveld$^{5,6}$
% ...
Texas Stroke Institute, Texas, United States of America.


\end{document}